\documentclass[12pt]{article} 
\usepackage{graphicx}
\newcommand{\be}{\begin{equation}}
\newcommand{\ee}{\end{equation}}
\newcommand{\bear}{\begin{eqnarray}}
\newcommand{\eear}{\end{eqnarray}}
\newcommand{\ba}{\begin{array}}
\newcommand{\ea}{\end{array}}

\newcommand{\CA}{{\cal A}} 
\newcommand{\CC}{{\cal C}} \newcommand{\CD}{{\cal D}}
 
\newcommand{\CN}{{\cal N}}

\newcommand{\CF}{{\cal F}} \newcommand{\CB}{{\cal B}}
\newcommand{\CV}{{\cal V}}  \newcommand{\CO}{{\cal O}}

\begin{document}

\begin{titlepage}
\vfill
\begin{flushright}
{\normalsize KIAS-P04012}\\
{\normalsize hep-th/0402115}\\
\end{flushright}

\vfill
\begin{center}
{\Large\bf A Note on AdS/CFT Dual of $SL(2,Z)$ Action on 3D
Conformal Field Theories with $U(1)$ Symmetry}

\vskip 0.3in

{ Ho-Ung Yee\footnote{\tt ho-ung.yee@kias.re.kr ,
tel:82-2-958-3730} }

\vskip 0.15in

 {\it School of Physics, Korea Institute for Advanced Study,} \\
{\it 207-43, Cheongryangri-Dong, Dongdaemun-Gu, Seoul 130-722,
Korea}
\\[0.3in]


\end{center}

\vfill

\begin{abstract}
\normalsize\noindent In this letter, we elaborate on the $SL(2,Z)$
action on three dimensional conformal field theories with $U(1)$
symmetry introduced by Witten, by trying to give an explicit
verification of the claim regarding holographic dual of the $S$
operation in AdS/CFT correspondence. A consistency check with the recently
proposed prescription on boundary condition of bulk fields when we deform the boundary CFT
in a non-standard manner is also discussed.
\end{abstract}

\vfill

\end{titlepage}
\setcounter{footnote}{0}

\baselineskip 18pt \pagebreak
\renewcommand{\thepage}{\arabic{page}}
\pagebreak

\section{Introduction }

Mirror symmetry found in three dimensional theories with extended
supersymmetry \cite{Intriligator:1996ex} gives us much insight
about non-trivial duality in quantum field theory. For the cases
with abelian gauge groups, it was shown \cite{Kapustin:1999ha}
that many aspects of duality may be derived by assuming a single
'elementary' duality, that is, the duality (in IR) between the
$\CN=4$ SQED with single flavor hypermultiplet and the free theory
of single hypermultiplet. The former has a global $U(1)$ symmetry
that shifts the dual photon scalar of $U(1)$ gauge field. This
symmetry is supposed to be the symmetry of $U(1)$ phase rotation
in the latter. Because magnetic vortices break the shift symmetry
of the dual photon, they can be identified to elementary
excitations in the free theory side.

Recently, it was observed in Ref.\cite{Witten:2003ya} that the
above simplest duality between vortex and particle may be seen as
an invariance under certain transformation on three-dimensional
CFT's. Specifically, given a CFT with global $U(1)$ symmetry, this
transformation is defined by gauging the $U(1)$ symmetry without
introducing gauge kinetic term. Although the above example is in
the context of supersymmetric version of this transformation,
there is no problem in defining this transformation in
non-supersymmetric cases, in general. The intriguing fact shown in
Ref.\cite{Witten:2003ya} is the possibility of extending this
transformation into a set of transformations forming the group
$SL(2,Z)$. The above transformation corresponds to $S$ with
$S^2=-1$, while the transformation $T$ with $(ST)^3=1$ was
introduced.

The meaning of this $SL(2,Z)$ in the space of 3D CFT's has been
studied in
Ref.\cite{Witten:2003ya,Burgess:2000kj,Leigh:2003ez,Zucchini:2003in}
for theories in which Gaussian approximation is valid in
calculating correlation functions.
These analysis identified the $SL(2,Z)$
as certain transformations of basic correlation functions of the
theory. While we may be almost convinced that the transformations
of correlation functions found in these analysis hold true in
general, its proof is currently limited to the theories with
Gaussian approximation.

As suggested in Ref.\cite{Witten:2003ya}, another way of
interpreting the $SL(2,Z)$ transformations may be provided by
AdS/CFT correspondence \cite{Maldacena:1997re}. According to
AdS/CFT, a global $U(1)$ symmetry in the CFT corresponds to having
a $U(1)$ gauge theory in the bulk, whose asymptotic value on the
boundary couples to the $U(1)$ current of the CFT. The $U(1)$
gauge theory in the bulk has a natural $SL(2,Z)$ duality
\cite{Witten:1995gf}. While it is easy to identify the $T$
operation in the CFT as the usual $2\pi$ shift of the bulk
$\theta$ parameter \cite{Witten:2003ya}, describing holographic
dual of the $S$ operation turns out to be much more subtle. It was
suggested that the $S$-transformed CFT is dual to the same gauge
theory in the bulk, but its $U(1)$ current couples to the
S-dualized gauge field. Note that the resulting CFT with different
coupling to the bulk field is not equivalent to the original CFT
\cite{Breitenlohner:jf,Klebanov:1999tb}.

Although a compelling discussion on holographic dual of the $S$
operation was provided in Ref.\cite{Witten:2003ya} using various
aspects of AdS/CFT \cite{Klebanov:1999tb,Witten:2001ua,Berkooz:2002ug}, and was further supported
in Ref.\cite{Zucchini:2003in} by explicitly calculating certain
correlation functions, a rigorous verification of the claim is
missing. In this letter, we propose a rigorous argument that fills
this gap.

\section{Setting up the stage}

This section is intended to give a brief review of relevant facts
in Ref.\cite{Witten:2003ya} on $SL(2,Z)$ transformations of 3D
CFT's, as a necessary preparation for the discussion in next
section.

A basic ingredient used in the discussion of
Ref.\cite{Witten:2003ya} is the equation, \be \int\CD
A\,\,\exp\Big(i\,I(A,B)\Big)\,\,=\,\,\int\CD
A\,\,\exp\Big(\frac{i}{2\pi}\int_Y d^3x\,\,
\epsilon^{ijk}A_i\partial_j
B_k\Big)\,\,=\,\,\delta(B)\quad,\label{basic}\ee where $A$ and $B$
are connections of line bundles on an oriented base three-manifold
$Y$. The delta function on the right-hand side means that $B$ is
zero, that is, its field strength vanishes and there is no
non-vanishing Wilson line. The path integral $\int \CD A$ in the
left-hand side includes summing over topologically distinct line
bundles as well as integral over trivial connections. For
topologically non-trivial connections, especially when a quantized
magnetic flux on a 2-dimensional cycle $\Sigma$ does not vanish,
\be \frac{1}{2\pi}\int_\Sigma F\,\,\neq\,\,0\quad,\ee it is not
possible to define a global connection $A$ such that $d A\,=\, F$.
In this case, we need to understand $I(A,B)$ as follows. Pick up a
compact-oriented four manifold $X$ whose boundary is $Y$, and
extend connections (and line bundles) $A$, $B$ on $Y$ to
connections $\CA$, $\CB$ on $X$. Then $I(A,B)$ is defined to be
\be \frac{1}{2\pi}\int_X\,\CF_\CA \wedge \CF_\CB\quad,\ee where
$\CF_\CA,\CF_\CB$ are the field strengths of $\CA,\CB$. Because
for any {\it closed} four manifold $\bar X$,
$\frac{1}{4\pi^2}\int_{\bar X}\,\,\CF_\CA \wedge \CF_\CB$ is an
integer Chern number, the above definition of $I(A,B)$ is easily
shown to be independent of extensions modulo $2\pi $. This is fine
as long as we are concerned only with $e^{i\, I(A,B)}$.

In Ref.\cite{Witten:2003ya}, several ways of showing (\ref{basic})
were given. In simple terms, we split $A\,=\,A_{triv}+A'$, where
$A_{triv}$ is a globally defined trivial connection, and $A'$ is a
representative of a given topologically non-trivial line bundle
(which does not have a global definition). Note that we can write
\be I(A,B)\,\,=\,\,\frac{1}{2\pi}\int_Y d^3x\,\,
\epsilon^{ijk}A_i^{triv}\partial_j
B_k\,\,+\,\,I(A',B)\quad,\label{split}\ee because
$\epsilon^{ijk}\partial_j B_k\,=\,\frac{1}{2}\epsilon^{ijk}F_{jk}$
is well-defined on $Y$. The path integral over $A_{triv}$ gives us
delta function setting the field strength of $B$ zero. Then, only
remaining component of $B$ is its possible Wilson line in
$H^1(Y,{\rm U}(1))$. With vanishing field strength of $B$, it is
clear as in (\ref{split}) that $I(A',B)$ is invariant under adding
trivial connection to $A'$, that is, $I(A',B)$ depends only on the
cohomology of the field strength of $A'$. This cohomology
(characteristic class) belongs to $H^2(Y,{\bf Z})$ as a
consequence of Dirac quantization (a Chern's theorem), or more
specifically, \be \frac{1}{2\pi}\int_\Sigma F\,\,\in {\bf
Z}\quad,\ee where $\Sigma$ is any integer coefficient 2-cycle.
Thus, we see that $I(A',B)$ is a kind of bilinear form, \be
H^2(Y,{\bf Z})\,\times\,H^1(Y,{\rm U}(1))\,\,\rightarrow\,\,{\rm
\bf R}\quad.\ee In Ref.\cite{Witten:2003ya}, this bilinear form
was identified and summing over $H^2(Y,{\bf Z})$ was shown to give
the remaining delta function setting Wilson line of $B$ zero. A
possible intuitive picture on this may be the following. Consider
a non-zero 1-cycle $\gamma$ on which there is a Wilson line
$e^{i\int_\gamma B}\in {\rm U}(1)$. We roughly consider $Y$ as a
product of $\gamma$ and two dimensional transverse space $\Sigma$,
and write $I(A',B)$ as \be I(A',B)\,\,\sim\,\,\frac{1}{2\pi}\int_Y
B\wedge
F_A\,\,\sim\,\,\frac{1}{2\pi}\int_\Sigma\,F_A\,\cdot\,\int_\gamma
B\,\sim\,n\,\cdot\,\int_{\gamma} B\quad, \ee where $n\,\in\,{\bf
Z}$. Hence, summing over $F_A\in H^2(Y,{\bf Z})$ involves
something like \be \sum_{n\in {\bf Z}}\,e^{i \,n \cdot\int_\gamma
B}\quad,\ee which imposes vanishing Wilson line, $e^{i\int_\gamma
B}=1$. This argument is intended to be just illustrative, and we
refer to Ref.\cite{Witten:2003ya} for rigorous derivation.

Now, we are ready to describe the $SL(2,Z)$ actions defined in
Ref.\cite{Witten:2003ya} on three dimensional conformal field
theories with global $U(1)$ symmetry. The definition of a
conformal field theory here means to specify the global $U(1)$
current $J^i$ and introduce a background gauge field $A_i$ without
kinetic term that couples to $J^i$. A theory is thus specified by
\be \bigg\langle \exp\Big(i\int_Y d^3x \,\,A_i
J^i\Big)\bigg\rangle\quad, \label{definition}\ee where
$\langle\ldots\rangle$ means to evaluate expectation value in the
given CFT. The above generating functional can produce all
correlation functions of $U(1)$ current $J^i$. The $S$ operation
is defined by letting $A_i$ be dynamical and introducing a
background gauge field $B_i$ with a coupling \be
I(A,B)=\frac{1}{2\pi}\int_Y d^3x\,\, \epsilon^{ijk}A_i\partial_j
B_k\quad, \ee that is, the transformed theory is now specified by
\be \int \CD A \,\,\bigg\langle \exp\Big(i\int_Y d^3x \,\,A_i
J^i\Big)\bigg\rangle \exp\Big(i \,I(A,B)\Big) \quad,
\label{transformed} \ee where $\langle \ldots \rangle$ means
expectation value in the original conformal field theory. Noting
that $ I(A,B)\,\sim\,\int_Y B\wedge F_A$, we see that the $U(1)$
current of the $S$-transformed theory  that $B$ couples is $\tilde
J^i=\frac{1}{2\pi}(\star
F_A)^i=\frac{1}{4\pi}\epsilon^{ijk}\,(F_A)_{jk}$. The $U(1)$
symmetry corresponding to this current is the shift symmetry of
dual photon scalar of $A_i$.

The definition of $T$ operation is a little subtle, because it
involves modifying a theory in a way which is not manifest in low
energy action that is supposed to define the theory. Concretely,
the $T$ operation is defined to shift the 2-point function of
$J^i$ by a contact term, \be \langle J^i(x)
J^j(y)\rangle\,\,\rightarrow\,\,\langle J^i(x)
J^j(y)\rangle\,\,+\,\,\frac{i}{2\pi}\epsilon^{ijk}\frac{\partial}{\partial
x^k}\delta^3(x-y)\quad.\label{modify}\ee Because the above contact
term has mass dimension 4, which is the right dimension of $JJ$
correlation, this term does not introduce any dimensionful
coupling. Moreover, it does not conflict with any symmetry of the
theory (in some cases \cite{Redlich:1983dv}, we need this term to
preserve gauge invariance). In fact, whenever there is freedom to
add local contact terms that are consistent with the symmetry of a
theory, this signals the intrinsic inability of our low energy
action in predicting them, and we have to {\it renormalize} them.
In other words, they must be treated as input parameters rather
than outputs. Note that this is not an unusual thing; it is an
essential concept of renormalization in quantum field theory. The
effect of the modification (\ref{modify}) on our generating
functional (\ref{definition}) is \be \bigg\langle \exp\Big(i\int_Y
d^3x \,\,A_i J^i\Big)\bigg\rangle\,\,\rightarrow\,\,\bigg\langle
\exp\Big(i\int_Y d^3x \,\,A_i J^i\Big)\bigg\rangle\,\cdot\,
\exp\Big(\frac{i}{4\pi}\int_Y d^3x\,\,\epsilon^{ijk}A_i\partial_j
A_k\Big)\quad,\ee which can be shown by first expanding the
exponent in series of $J$ and re-exponentiating the effects of $T$
operation on $J$ correlation functions.

Another fact in Ref.\cite{Witten:2003ya}, which is needed to show
the $SL(2,Z)$ group structure of the above transformations is, \be
\int\CD\CA\,\,\exp\bigg(i\,I(A)\bigg)\,\,=\,\,1\quad,\label{trivial}\ee
up to possible phase factor \cite{Witten:1995gf,Vafa:1994tf}. This
equation should be understood as a statement that the theory has
only one physical state and trivial \cite{Elitzur:1989nr}. Here,
\be I(A)\,\,=\,\,\frac{1}{4\pi}\int_Y
d^3x\,\,\epsilon^{ijk}A_i\partial_j A_k
\,\,\equiv\,\,\frac{1}{16\pi}\int_X
d^4x\,\,\epsilon^{ijkl}F_{ij}F_{kl}\,\,=\,\,\frac{1}{4\pi}\int_X
F\wedge F \quad,\ee defined with some extension over $X$ similarly
as before \cite{Deser:1981wh}. This is well-defined modulo $2\pi$
for a spin manifold $Y$.
Using (\ref{basic}) and (\ref{trivial}), it is
readily shown that $S$ and $T$ satisfy the $SL(2,Z)$ generating
algebra, $(ST)^3=1$ and $S^2=-1$, where $-1$ is the transformation
$J^i\rightarrow -J^i$ commuting with everything.

\section{Holographic dual of the S operation in AdS/CFT}

We now try to elaborate on the claim in Ref.\cite{Witten:2003ya}
and to give an explicit proof that the $S$ operation on CFT's is
dual to the abelian S-duality in the bulk AdS in AdS/CFT
correspondence.

Let $X$ denote the bulk AdS, and $\partial X =Y$ be our
space-time. Let $\CA$ be the $U(1)$ gauge field in the bulk whose
boundary value couples to the global $U(1)$ current $J^i$ in the
CFT side. According to AdS/CFT, we have \be \bigg\langle
\exp\Big(i\int_Y d^3x \,\,A_i J^i\Big)\bigg\rangle\,\,=
\,\,\int_{\CA_i\rightarrow A_i} \CD \CA\,\, \exp\Big(i
S(\CA)\Big)\quad, \label{adscft}\ee where
$S(\CA)=\frac{1}{e^2}\int_X \CF_\CA\wedge\ast\CF_\CA+\cdots$ is
the action of the bulk gauge field and we omitted other bulk
fields for simplicity. Before considering holographic dual of $S$
operation, it is easy to identify from (\ref{adscft}) the
holographic dual of $T$ operation as in Ref.\cite{Witten:2003ya}.
The $T$ operation simply multiplies $e^{i\,I(A)}$ in both sides of
(\ref{adscft}). But, note that $I(A)=\frac{1}{4\pi}\int_X
\CF_\CA\wedge\CF_\CA$ modulo $2\pi$ irrespective of the bulk
extension $\CA$ as long as its boundary value is fixed, hence in
the right-hand side, multiplying $e^{i\,I(A)}$ is equivalent to
shifting the bulk $\theta$ term, \be S(\CA)\,\, \supset
\,\,\frac{\theta}{8\pi^2}\int_X \CF_\CA\wedge\CF_\CA \quad,\ee by
$\theta\rightarrow\theta+2\pi$.

Now, using (\ref{adscft}), we want to show that
(\ref{transformed}) is nothing but the bulk path integral of the
same bulk theory, but with the boundary condition that the 'dual'
field $\CV$ has the specified boundary value $B_i$. In terms of
the original field $\CA$, this corresponds to specifying electric
field on the boundary, instead of specifying magnetic field. (When
$B_i=0$, the boundary condition in terms of $\CA$ is that the
electric field vanishes on the boundary, as given in
Ref\cite{Witten:2003ya}.)

 Using AdS/CFT
and the fact that $I(A,B)=\frac{1}{2\pi}\int_Y d^3x\,\,
\epsilon^{ijk}A_i\partial_j B_k$ can be written as a bulk integral
(up to mod $2\pi$) \be I(A,B)\,\,=\,\,\frac{1}{2\pi}\int_X
\CF_\CB\wedge\CF_\CA\quad, \ee where $\CB$ and $\CA$ are
'arbitrary' extensions of $B_i$ and $A_i$, we have \bear
&&\bigg\langle \exp\Big(i\int_Y d^3x \,\,A_i J^i\Big)\bigg\rangle
\exp\Big(\frac{i}{2\pi}\int_Y d^3x\,\, \epsilon^{ijk}A_i\partial_j B_k\Big)\nonumber\\
&=&
\int_{\CA_i\rightarrow A_i} \CD \CA\,\,
\exp\Big(i S(\CA)+\frac{i}{2\pi}\int_X \CF_\CB\wedge\CF_\CA\Big)\quad,
\label{core}
\eear
where $\CB$ is some fixed extension of $B_i$.
(\ref{transformed}) is the integral of this quantity over the boundary value $A_i$, hence (\ref{transformed})
is equal to the RHS of (\ref{core}) without any boundary conditions on $\CA$,
\bear
&&\int \CD A \,\,\bigg\langle \exp\Big(i\int d^3x \,\,A_i J^i\Big)\bigg\rangle
\exp\Big(\frac{i}{2\pi}\int d^3x\,\, \epsilon^{ijk}A_i\partial_j B_k\Big)\nonumber\\
&=&
\int \CD \CA\,\,
\exp\Big(i S(\CA)+\frac{i}{2\pi}\int_X \CF_\CB\wedge\CF_\CA\Big)\quad.\label{intermed}
\eear

\begin{figure}[t]
\begin{center}
\scalebox{1}[1]{\includegraphics{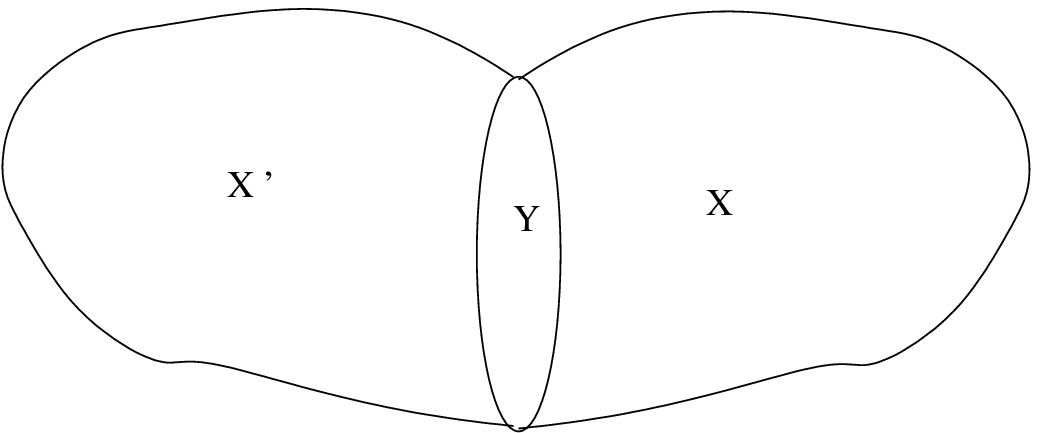}}
\par
\vskip-2.0cm{}
\end{center}
\caption{\small} \label{domain}
\end{figure}

We now perform a dualizing procedure in the bulk $X$, which is
similar to the one in Ref.\cite{Witten:1995gf}, but appropriately
taking care of the fact that our space-time now has a boundary
$\partial X=Y$. First we want to argue that, for a bulk 2-form
field $G$, the integral \be \int_{\CV\rightarrow 0}\CD \CV \,\,
\exp\Big(\frac{i}{2\pi}\int_X \CF_\CV \wedge G\Big) \label{delta}
\ee over all possible connections $\CV$ (and also sum over line
bundles) in $X$ with boundary condition that $\CV$ vanishes on $Y$
(up to gauge transformations), gives a delta function on $G$ that
precisely says $G$ is a field strength of some connection of a
line bundle. To show this, we consider a "closed" 4-manifold $\bar
X$ which is obtained from $X$ by attaching on $\partial X=Y$ a
orientation reversed copy of $X$ which we call $X'$, as in
Fig.\ref{domain}. We also consider a 2-form field $\bar G$ on
$\bar X$, whose value on $X'$ is the identical copy of $G$ on $X$.
It is clear that $G$ is a field strength of some connection on $X$
if and only if $\bar G$ is a field strength of some connection on
$\bar X$. As $\bar X$ is closed, we can use the well-known
procedure of requiring $\bar G$ to be a field strength
\cite{Witten:1995gf}; the integral \be \int \CD \bar\CV \,\,
\exp\Big(\frac{i}{2\pi}\int_{\bar X} \CF_{\bar\CV} \wedge \bar
G\Big) \ee over connections $\bar \CV$ on $\bar X$ gives a delta
function imposing that $\bar G$ is a field strength of some
connection on $\bar X$. Simply put, the integration over trivial
part in $\bar\CV$ imposes that $\bar G$ be a closed 2-form, while
the remaining sum over line bundles requires $\bar G$ to satisfy
Dirac quantization, $\bar G\,\in\,H^2(\bar X,{\bf Z})$.

Thus, when expressed in terms of $G$, it gives the desired delta
function (up to a constant factor) that says $G$ should be a field
strength on $X$. Now, we can split $\bar\CV$ on $\bar X$ into a
connection $\CV$ on $X$ and a connection $\CV'$ on $X'$, and we
have \bear \frac{i}{2\pi}\int_{\bar X} \CF_{\bar\CV} \wedge \bar G
&=&\frac{i}{2\pi}\int_X \CF_\CV \wedge G+\frac{i}{2\pi}\int_{X'} \CF_{\CV'}\wedge G\nonumber\\
&=&\frac{i}{2\pi}\int_X \CF_\CV \wedge G-\frac{i}{2\pi}\int_{X} \CF_{\CV'}\wedge G\nonumber\\
&=&\frac{i}{2\pi}\int_X (\CF_\CV-\CF_{\CV\,'}) \wedge G \quad,
\eear
where in the last line, we consider $\CV\,'$ as a connection on $X$, but with the minus sign
in the integral due to orientation reversal. Note that $\CV$ and $\CV'$ should agree on the boundary $Y$,
as they are from a common $\bar \CV$ on $\bar X$, hence we can rewrite the path integral over $\bar\CV$
into
\be
\int \CD \bar\CV=\int_{(\CV-\CV')\rightarrow 0} \CD \CV \,\,\CD\CV\,'
\ee
From the above two observations, we have
\bear
&&\int \CD \bar\CV \,\,
\exp\Big(\frac{i}{2\pi}\int_{\bar X} \CF_{\bar\CV} \wedge \bar G\Big)\nonumber\\
&=&
\int_{(\CV-\CV')\rightarrow 0} \CD \CV \,\,\CD\CV\,'
\exp\Big(\frac{i}{2\pi}\int_X (\CF_\CV-\CF_{\CV\,'}) \wedge G \Big)\nonumber\\
&=& \bigg[\int \CD\CV\,'\bigg]\cdot
\int_{\CV\rightarrow 0}\CD \CV \,\,
\exp\Big(\frac{i}{2\pi}\int_X \CF_\CV \wedge G\Big)\quad,
\eear
where we have changed the variable $(\CV-\CV\,')\rightarrow \CV$ in the last line.
Thus, (\ref{delta}) indeed gives a desired delta function (up to a constant factor).

Now, we are ready to perform the duality procedure in a space-time
with boundary. Introduce a 2-form field $G$ and replace every
$\CF_\CA$ in the action with $\CF_\CA+G$. Also introduce a
connection $\CV$ with the boundary condition that $\CV$ vanishes
on $Y$, and add the coupling \be \frac{i}{2\pi}\int_X \CF_\CV
\wedge G\quad.\label{coupling} \ee The resulting action is
invariant under the extended gauge transform, \be
\CA\,\,\rightarrow\,\,\CA+\CC\quad,\quad
G\,\,\rightarrow\,\,G-\CF_\CC\quad, \label{gauge} \ee where $\CC$
is an arbitrary connection in $X$. Precisely because $\CV$
vanishes on $Y$, (\ref{coupling}) is invariant under (\ref{gauge})
modulo $2\pi i$. Let us explain this fact in some detail. The
vanishing connection on $Y$ can be extended to a trivial (globally
defined one form) connection on $X$, say $\CV'$. We also know that
\be \frac{i}{2\pi}\int_X \CF_{\CV'}\wedge
\CF_\CC\,\,=\,\,\frac{i}{2\pi}\int_X \CF_{\CV}\wedge
\CF_\CC\quad,\ee modulo $2\pi i$ because $\CV'$ and $\CV$ agree on
$Y$. Being trivial, $\CF_{\CV'}$ can be written as
$\CF_{\CV'}=d\CV'$ globally on $X$, and performing partial
integration, we have \be \frac{1}{2\pi}\int_X \CF_{\CV'}\wedge
\CF_\CC\,\,=\,\,\frac{1}{2\pi}\int_Y \CV'\wedge
\CF_\CC\,\,=\,\,0\quad,\ee because $\CV'$ vanishes on $Y$.

 We then consider $G$ and $\CV$ as dynamical, and mod out the theory
with gauge equivalence. If we integrate over $\CV$ first, it gives
a constraint that $G$ is a field strength of some connection $C$
by the discussion in the previous paragraphs. Then, by gauge
fixing, we can set $G=0$ and recover the original theory of $\CA$.
The equivalent dual theory in terms of $\CV$ is obtained by first
gauge fixing $\CA=0$, and integrating over $G$. Applying this to
(\ref{intermed}), we get \bear &&\int \CD \CA\,\,
\exp\Big(i S(\CF_\CA)+\frac{i}{2\pi}\int_X \CF_\CB\wedge\CF_\CA\Big)\nonumber\\
&=&
\int_{\CV\rightarrow 0} \CD\CV\,\int \CD G\,\,
\exp\Big(i S(G)+\frac{i}{2\pi}\int_X (\CF_\CB+\CF_\CV)\wedge G\Big)\nonumber\\
&=& \int_{\CV\rightarrow B_i} \CD\CV\,\int \CD G\,\, \exp\Big(i
S(G)+\frac{i}{2\pi}\int_X \CF_\CV\wedge G\Big)\nonumber\\
&=& \int_{\CV\rightarrow B_i} \CD\CV \,\, \exp\Big(i
S_D(\CF_\CV)\Big) \quad, \label{dual}\eear
where
in the third line, we changed the variable $\CB+\CV\rightarrow\CV$
with the new boundary condition that $\CV$ goes to the specified
$B_i$ on $Y$. In the last line, integrating over $G$ gives the dual bulk
action $S_D(\CF_\CV)$ in terms of the dual gauge field $\CV$ with the coupling constant $-\frac{1}{\tau}$,
and we have the
desired boundary condition for $\CV$ on $Y$.

At this point, it would be clarifying to see explicitly
the relation between the boundary condition for the dual field $\CV$
that we derived above, and the boundary condition in terms of the original field $\CA$ \cite{Witten:2003ya}.
In the bulk AdS,
the dual field $\CV$ is nothing but a non-local change of variable from the
original variable $\CA$. In the case of vanishing $\theta$ angle
\footnote{This is just for simplicity. The case with non-vanishing $\theta$ angle is
similar \cite{Witten:1995gf}.},
they are related by
\be
(\CF_\CA)_{\mu\nu}\,\,=\,\,\frac{e^2}{8\pi}\,\epsilon_{\mu\nu\alpha\beta}\,(\CF_\CV)^{\alpha\beta}\quad.
\label{relation}
\ee
Now, in Poincare coordinate $(x_0,\vec{x})$, (with the boundary at $x_0=0$)
\be
ds^2=\frac{dx_0^2+d\vec{x}^2}{x_0^2}\quad,
\ee
the usual boundary condition specifying the value of gauge field on the boundary corresponds to
specifying the "magnetic" component $M_i=\frac{1}{2}\epsilon_{ijk}F^{jk}$, $i,j,k=1,2,3$.
Note that the gauge invariant data of the boundary value of gauge field is $M_i$.
Because (\ref{relation}) interchanges the "magnetic" component $M_i$
with the "electric" component of $\CA$, $E_i=\partial_0 \CA_i$ (in the gauge $A_0=0$), we see that
in terms of the original field $\CA$, the $S$-transformed CFT is mapped to the bulk AdS theory with
"electric" boundary condition. Note that "magnetic" and "electric" boundary conditions are natural
counterparts of Dirichlet and Neumann boundary conditions for scalar field, and they are naturally
expected to be conjugate with each other in AdS/CFT. We will come to this point in the next section.
In fact, we need to look at the $T$-transformation more carefully in this respect.
The AdS dual of the $T$-transformation of 3D CFT was identified as a $2\pi$ shift of the bulk $\theta$-angle, while
the "magnetic" boundary condition is unchanged. In the presence of $\theta$-angle, the 'electric' component
naturally conjugate to the 'magnetic' component (or more precisely, the value $A_i$ on the boundary)
has a term proportional to $\theta$-angle. This is most easily
seen from the fact that the natural conjugate variable to $A_i$ is obtained by varying the action
w.r.t. $\partial_0 \CA_i$.\footnote{If we consider $x_0$ as a time variable, this is the Witten effect. This should be true
even in Euclidean case when considering 'naturally' conjugate boundary variables on $x_0=0$.}
Denoting this as $D_i$, we have
\be
D_i\,\,=\,\,\frac{1}{e^2}\partial_0\CA_i\,+\,\frac{\theta}{8\pi^2}M_i\quad,
\ee
and shifting $\theta$ results in shifting of $D_i$ by a unit of 'magnetic' component $M_i$.
Note that $\partial^i D_i=0$ due to Bianchi identity of $M_i$ and the equation of motion $\partial^i\partial_0 \CA_i=0$.

\section{In view of boundary deformations}

In the last section, we observed that the AdS dual of $S$-operation on 3D CFT interchanges
the 'magnetic' and 'electric' boundary conditions, while $T$-operation corresponds to shifting
the 'electric' component $D_i$ by a unit of 'magnetic' component. Though $T$-operation doesn't
really change the boundary condition by itself, it has a non-trivial effect when combined with $S$.
With appropriate normalization, we can represent the $S$ and $T$ action on boundary conditions as
\bear
S\,\,&:&\,\,\left(\ba{c}
D_i\\M_i\ea\right)\rightarrow \left(\ba{cc}0&-1\\1&0\ea\right)\left(\ba{c}
D_i\\M_i\ea\right)\nonumber\\
T\,\,&:&\,\,\left(\ba{c}
D_i\\M_i\ea\right)\rightarrow \left(\ba{cc}1&1\\0&1\ea\right)\left(\ba{c}
D_i\\M_i\ea\right)\quad,
\eear
where we take the usual 'magnetic' boundary condition in terms of transformed variable.
This gives a natural correspondence between the $SL(2,Z)$ action on 3D CFT's with
the $SL(2,Z)$ action on boundary condition (or bulk gauge field).

In this section, we give another concrete evidence of this picture in the context of
the recently proposed prescription \cite{Witten:2001ua} on boundary conditions when we
deform the boundary CFT in a non-standard manner.
The proposed prescription in Ref.\cite{Witten:2001ua} is for scalar fields in the bulk, and it goes as follows.
Suppose we have a scalar field $\phi$ in the bulk, whose asymptotic behavior near the boundary $x_0=0$ is
\be
\phi(x_0,\vec{x})\,\,\sim\,\,A(\vec{x})x_0^{\Delta_-}+B(\vec{x})x_0^{\Delta_+}\quad.
\ee
We consider the CFT on $x_0=0$ defined by the boundary condition $A(\vec{x})=0$.
By standard AdS/CFT dictionary, $A(\vec{x})$ couples to a scalar operator $\CO$ of dimension $\Delta_+$ on the boundary.
The expectation value of $\CO$ in this deformed CFT is given by $B(\vec{x})$,
\be
\langle \CO(\vec{x})\rangle_{A}\,\,\sim\,\, B(\vec{x})\quad.
\ee
They are natural conjugate pair of source and expectation value.
The question is what would be the boundary condition when we deform the boundary CFT (defined by $A(\vec{x})=0$)
in a more general manner,
\be
S_{\rm CFT}\,\,\rightarrow\,\,S_{\rm CFT}+W(\CO)\quad,
\ee
where $W$ is an arbitrary (possibly non-local) function of $\CO(\vec{x})$.
The proposal in Ref.\cite{Witten:2001ua} is to take the following boundary condition on $A(\vec{x})$ and $B(\vec{x})$,
\be
A(\vec{x})\,\,=\,\,\frac{\delta W(\CO)}{\delta \CO(\vec{x})}\bigg\arrowvert_{\CO(\vec{x})\rightarrow B(\vec{x})}\quad.
\label{proposalboundary}
\ee

The situation with bulk gauge field in ${\rm AdS}_4/{\rm CFT}_3$ correspondence looks similar to the case
of bulk scalar field with its mass such that two CFT's are possible.
We have two naturally conjugate variables $(A_i,D_i)$ in the boundary, and two different boundary conditions
are possible.
Hence, the proposal (\ref{proposalboundary}) can be naturally extended to include the case of gauge fields,
and we will show that this extension indeed reproduces the results in the previous sections, providing
a compelling check for the proposal applied to gauge fields.
We start with the CFT defined by the usual 'magnetic' boundary condition specifying gauge field components
tangential to the boundary as $x_0\rightarrow 0$,
\be
\CA_i\,\,\rightarrow A_i\,,\,\,i=1,2,3\quad.
\ee
As usual, this corresponds to deforming the CFT by adding the coupling,
\be
\delta S_{\rm CFT}\,\,=\,\, \int d^3x\,A_i\,J^i\quad,\label{linearone}
\ee
where $J^i$ is the 3D $U(1)$ current. Now, we ask the question of what boundary condition we
take when we deform the CFT with an arbitrary function of $J^i(\vec{x})$,
\be
\delta S_{\rm CFT}\,\,=\,\,W[J^i]\quad,
\ee
instead of a linear one (\ref{linearone}). Recalling that
\be
D^i(\vec{x})=\frac{\delta S_{\rm bulk}}{\delta \partial_0 \CA_i(\vec{x})}
\ee
is the "electric" field  that is canonically conjugate to $\CA_i(\vec{x})$,
a direct analogy with the case of scalar fields (\ref{proposalboundary})
suggests the following prescription on the boundary condition as $x_0\rightarrow 0$,
\be
A_i(\vec{x})\,\,=\,\,\frac{\delta W[J]}{\delta J^i(\vec{x})}\bigg\arrowvert_{J^i(\vec{x})\rightarrow D^i(\vec{x})}\quad,
\label{suggestion}
\ee
where $\CA_i\rightarrow A_i$.

Having this in mind, let us go back to our $SL(2,Z)$ actions on 3D CFT and consider the action
given by $ST^n$, which corresponds to the matrix,
\be
\left(\ba{cc}0&-1\\1&0\ea\right)\left(\ba{cc}1&n\\0&1\ea\right)=\left(\ba{cc}0&-1\\1&n\ea\right)\quad.
\ee
By definition, the partition function of the transformed CFT is given by
\be
Z_{\rm after}\,\,=\,\,\int \CD B_i\, \left<\exp(i\int d^3x\,B_i J^i+n\cdot\frac{i}{4\pi}\int d^3x\,\epsilon^{ijk}B_i\partial_j B_k)
\right>\quad,\label{partition}
\ee
where $<\ldots>$ and $J^i$ are expectation values and $U(1)$ current, respectively, of the original CFT,
and $B_i$ is the intermediate connection variable in defining the $S$ operation. Because the exponent is quadratic
in $B_i$, we can perform the path integral over $B_i$ explicitly. We introduce the gauge fixing term
$i\int d^3x\, \xi(\partial_i B_i)^2$, and the $B_i$ propagator is
\be
S_{ij}(p)\,\,\equiv\,\,\left<B_i(p)B_j(-p)\right>\,\,=\,\,i\frac{p_i p_j}{2\xi(p^2)^2}+\frac{2\pi }{n p^2}\,\epsilon_{ijk}\,p^k\quad.
\ee
Using this, (\ref{partition}) is given by
\bear
Z_{\rm after}&=&\left<\exp\bigg(-\frac{1}{2}\int \frac{d^3p}{(2\pi)^3}J^i(-p)S_{ij}(p)J^j(p)\bigg)\right>\nonumber\\
&=&\left<\exp\bigg(-\frac{\pi}{n}\int\frac{d^3p}{(2\pi)^3}J^i(-p)\frac{\epsilon_{ijk}p^k}{p^2}J^j(p)\bigg)\right>\quad,
\eear
where $J^i(p)\equiv \int d^3x\,e^{-ipx}J^i(x)$, and we have used the Ward identity $\left<p_i J^i(p)\ldots\right>=0$.
Now, looking at the last expression, it is clear that the transformed theory is nothing but the original CFT with the
deformation $W[J]$ given by
\bear
\delta S_{\rm CFT}&=&W[J]\,\,=\,\,\frac{i\pi}{n}\int\frac{d^3p}{(2\pi)^3}J^i(-p)\frac{\epsilon_{ijk}p^k}{p^2}J^j(p)\nonumber\\
&=& \frac{i\pi}{n}\int\frac{d^3p}{(2\pi)^3}\int d^3x\int d^3y\,e^{ipx}e^{-ipy}\frac{\epsilon_{ijk}p^k}{p^2} J^i(x)J^j(y)\quad.
\eear
Hence, according to our proposal (\ref{suggestion}), the bulk AdS gauge theory of the transformed CFT has a modified boundary
condition,
\be
A_i(x)=\frac{2\pi i}{n}\int\frac{d^3p}{(2\pi)^3}\int d^3y\, e^{ipx}e^{-ipy}\frac{\epsilon_{ijk}p^k}{p^2}J^j(y)
\Bigg\arrowvert_{J^j(y)\rightarrow D^j(y)}\quad.
\ee
To see clearly what this means, take the $x$-derivative, $\frac{1}{2}\epsilon^{mni}\partial_n$, on both sides.
The left-hand side gives
the "magnetic" component $M^m(x)=\frac{1}{2}\epsilon^{mni}\partial_n A_i(x)$, while the right-hand side becomes
\bear
&&-\frac{\pi}{n}\int\frac{d^3p}{(2\pi)^3}\int d^3y\,e^{ipx}e^{-ipy}\epsilon^{mni}\epsilon_{ijk}
\frac{p_n p^k}{p^2} J^j(y)\Bigg\arrowvert_{J^j(y)\rightarrow D^j(y)}\nonumber\\
&=&-\frac{\pi}{n}\int\frac{d^3p}{(2\pi)^3}\int d^3y\,e^{ipx}e^{-ipy}(\delta^m_j -\frac{p^m p_j}{p^2})
 J^j(y)\Bigg\arrowvert_{J^j(y)\rightarrow D^j(y)}\nonumber\\
&=&-\frac{\pi}{n}\int\frac{d^3p}{(2\pi)^3}\int d^3y\,e^{ipx}e^{-ipy}
 J^m(y)\Bigg\arrowvert_{J^j(y)\rightarrow D^j(y)}\nonumber\\
 &=& -\frac{\pi}{n}
 J^m(x)\Bigg\arrowvert_{J^j(x)\rightarrow D^j(x)}\,\,=\,\,-\frac{\pi}{n}\,D^m(x)\quad,
\eear
where in going from the second line to the third, we again used the Ward identity for $J^i$.

In summary, the AdS bulk gauge field for the transformed CFT has the boundary condition;
$n\cdot M_i(\vec{x})+ D_i(\vec{x})=0$ (with appropriate normalization absorbing $\pi$).
Observe that this matches precisely with the result of the previous sections, because $ST^n$
corresponds to performing first the change
\be
\left(\ba{c}D_i\\M_i\ea\right)\,\,\rightarrow\,\,\left(\ba{cc}0&-1 \\ 1&n\ea\right)\left(\ba{c}D_i\\M_i\ea\right)
\,\,=\,\,\left(\ba{c}-M_i\\n\cdot M_i+ D_i\ea\right)\quad,
\ee
before taking the usual "magnetic" boundary condition $M_i=0$.

\vskip 1cm \centerline{\large \bf Acknowledgement} \vskip 0.5cm We
would like to thank Edward Witten for helpful comments, Kimyeong
Lee, Kyungho Oh, Jae-Suk Park and Piljin Yi for helpful
discussions. This work is partly supported by grant
No.R01-2003-000-10391-0 from the Basic Research Program of the
Korea Science \& Engineering Foundation.

 \vfil


\begin{thebibliography}{99} \frenchspacing

\bibitem{Witten:2003ya}
E.~Witten, ``SL(2,Z) action on three-dimensional conformal field
theories with Abelian symmetry,'' [arXiv:hep-th/0307041].


\bibitem{Witten:1995gf}
E.~Witten,
``On S duality in Abelian gauge theory,''
Selecta Math.\  {\bf 1}, 383 (1995)
[arXiv:hep-th/9505186].

\bibitem{Vafa:1994tf}
C.~Vafa and E.~Witten, ``A strong coupling test of S duality,''
Nucl.\ Phys.\ B {\bf 431}, 3 (1994) [arXiv:hep-th/9408074].


\bibitem{Burgess:2000kj}
C.~P.~Burgess and B.~P.~Dolan,
 ``Particle-vortex duality and the modular group: Applications to the  quantum
Hall effect and other 2-D systems,'' [arXiv:hep-th/0010246].


\bibitem{Leigh:2003ez}
R.~G.~Leigh and A.~C.~Petkou,
``SL(2,Z) action on
three-dimensional CFTs and holography,'' JHEP {\bf 0312}, 020
(2003) [arXiv:hep-th/0309177].

\bibitem{Zucchini:2003in}
R.~Zucchini,
 ``Four dimensional Abelian duality and SL(2,Z) action in three dimensional
conformal field theory,'' [arXiv:hep-th/0311143].


\bibitem{Intriligator:1996ex}
K.~A.~Intriligator and N.~Seiberg,
``Mirror symmetry in three
dimensional gauge theories,'' Phys.\ Lett.\ B {\bf 387}, 513
(1996) [arXiv:hep-th/9607207].


\bibitem{Kapustin:1999ha}
A.~Kapustin and M.~J.~Strassler,
``On mirror symmetry in three
dimensional Abelian gauge theories,'' JHEP {\bf 9904}, 021 (1999)
[arXiv:hep-th/9902033].

\bibitem{Maldacena:1997re}
J.~M.~Maldacena,
``The large N limit of superconformal field
theories and supergravity,'' Adv.\ Theor.\ Math.\ Phys.\  {\bf 2},
231 (1998) [Int.\ J.\ Theor.\ Phys.\  {\bf 38}, 1113 (1999)]
[arXiv:hep-th/9711200].



\bibitem{Shapere:1988zv}
A.~D.~Shapere and F.~Wilczek, ``Self-dual models with theta
terms,'' Nucl.\ Phys.\ B {\bf 320}, 669 (1989).\\
Y.~Kim and K.~M.~Lee, ``Vortex dynamics in selfdual Chern-Simons
Higgs systems,'' Phys.\ Rev.\ D {\bf 49}, 2041 (1994)
[arXiv:hep-th/9211035].\\
S.~J.~Rey and A.~Zee, ``Self-duality of three-dimensional
Chern-Simons theory,'' Nucl.\ Phys.\ B {\bf 352}, 897 (1991).\\
C.A.Lutken and G.G.Ross, ``Duality in the quantum Hall system,''
Phys.Rev.{\bf B45}(1992) 11837, Phys.Rev.{\bf
B48}(1993)2500.\\
D.-H.Lee, S.Kivelson, and S.-C.Zhang, Phys.Lett.{\bf 68}(1992)
2386, Phys.Rev.{\bf B46} (1992) 2223\\
C.~A.~Lutken,
 ``Geometry of renormalization group flows constrained by discrete global
symmetries,'' Nucl.\ Phys.\ B {\bf 396}, 670 (1993).\\
B.~P.~Dolan, ``Duality and the Modular Group in the Quantum Hall
Effect,'' J.\ Phys.\ A {\bf 32}, L243 (1999)
[arXiv:cond-mat/9805171].\\
C.~P.~Burgess, R.~Dib and B.~P.~Dolan,
``Derivation of the
Semi-circle Law from the Law of Corresponding States,'' Phys.\
Rev.\ B {\bf 62}, 15359 (2000) [arXiv:cond-mat/9911476].

\bibitem{Deser:1981wh}
S.~Deser, R.~Jackiw and S.~Templeton, ``Topologically massive
gauge theories,'' Annals Phys.\  {\bf 140}, 372 (1982)
[Erratum-ibid.\ {\bf 185}, 406.1988\ APNYA,281,409 (1988\
APNYA,281,409-449.2000)].


\bibitem{Breitenlohner:jf}
P.~Breitenlohner and D.~Z.~Freedman, ``Stability in gauged
extended supergravity,'' Annals Phys.\  {\bf 144}, 249 (1982).

\bibitem{Klebanov:1999tb}
I.~R.~Klebanov and E.~Witten, ``AdS/CFT correspondence and
symmetry breaking,'' Nucl.\ Phys.\ B {\bf 556}, 89 (1999)
[arXiv:hep-th/9905104].


\bibitem{Elitzur:1989nr}
S.~Elitzur, G.~W.~Moore, A.~Schwimmer and N.~Seiberg, ``Remarks On
The Canonical Quantization Of The Chern-Simons-Witten Theory,''
Nucl.\ Phys.\ B {\bf 326}, 108 (1989).

\bibitem{Redlich:1983dv}
A.~N.~Redlich,
 ``Parity violation and gauge noninvariance of the effective gauge field action
in three-dimensions,'' Phys.\ Rev.\ D {\bf 29}, 2366 (1984).

\bibitem{Witten:2001ua}
E.~Witten, ``Multi-trace operators, boundary conditions, and
AdS/CFT correspondence,'' arXiv:hep-th/0112258.


\bibitem{Berkooz:2002ug}
M.~Berkooz, A.~Sever and A.~Shomer,
 ``Double-trace deformations, boundary conditions and spacetime
singularities,'' JHEP {\bf 0205}, 034 (2002)
[arXiv:hep-th/0112264].\\
P.~Minces, ``Multi-trace operators and the generalized AdS/CFT
prescription,'' Phys.\ Rev.\ D {\bf 68}, 024027 (2003)
[arXiv:hep-th/0201172].\\
O.~Aharony, M.~Berkooz and E.~Silverstein, ``Multiple-trace
operators and non-local string theories,'' JHEP {\bf 0108}, 006
(2001) [arXiv:hep-th/0105309].\\
V.~K.~Dobrev, ``Intertwining operator realization of the AdS/CFT
correspondence,'' Nucl.\ Phys.\ B {\bf 553}, 559 (1999)
[arXiv:hep-th/9812194].\\
I.~R.~Klebanov, ``Touching random surfaces and Liouville
gravity,'' Phys.\ Rev.\ D {\bf 51}, 1836 (1995)
[arXiv:hep-th/9407167].\\
I.~R.~Klebanov and A.~Hashimoto, ``Nonperturbative solution of
matrix models modified by trace squared terms,'' Nucl.\ Phys.\ B
{\bf 434}, 264 (1995) [arXiv:hep-th/9409064].\\
S.~S.~Gubser and I.~Mitra, ``Double-trace operators and one-loop
vacuum energy in AdS/CFT,'' Phys.\ Rev.\ D {\bf 67}, 064018 (2003)
[arXiv:hep-th/0210093].\\
S.~S.~Gubser and I.~R.~Klebanov,
 ``A universal result on central charges in the presence of double-trace
deformations,'' Nucl.\ Phys.\ B {\bf 656}, 23 (2003)
[arXiv:hep-th/0212138].













\end{thebibliography}
\end{document}